\documentclass{pasa}%

\usepackage{graphicx}
\usepackage{slashed}
\usepackage{subfigure}
\title[Fermionic Light Dark Matter particles and the New Physics of Neutron Stars]{Fermionic Light Dark Matter particles and the New Physics of Neutron Stars}

\author[M. Cerme\~no et al.]{M. Cerme\~no$^{1,}$\thanks{marinacgavilan@usal.es}, M. \'Angeles P\'erez-Garc\'ia$^{1,}$\thanks{mperezga@usal.es} and Joseph Silk$^{2, 3, 4,}$\thanks{silk@iap.fr}
\affil{$^1$Department of Fundamental Physics, University of Salamanca, Plaza de la Merced s/n 37008 Spain}%
\affil{$^2$Institut d'Astrophysique,  UMR 7095 CNRS, Universit\'e Pierre et Marie Curie, 98bis Blvd Arago, 75014 Paris, France\\ $^3$Department of Physics and Astronomy, The Johns Hopkins University, Homewood Campus, Baltimore MD 21218, USA\\
$^4$Beecroft Institute of Particle Astrophysics and Cosmology, Department of Physics, University of Oxford, Oxford OX1 3RH, UK}
}%

\jid{PASA}
\doi{10.1017/pas.\the\year.xxx}
\jyear{\the\year}

\usepackage{aas_macros}
\usepackage{hyperref} 
\hypersetup{colorlinks,citecolor=blue,linkcolor=blue,urlcolor=blue}

\hypersetup{draft}

\begin{document}

\begin{frontmatter}
\maketitle

\begin{abstract}
 Dark Matter constitutes most of the matter in the presently accepted cosmological model for our Universe. The extreme conditions of ordinary baryonic matter, namely high density and compactness, in Neutron Stars make these objects suitable to gravitationally accrete such a massive component provided interaction strength between both, luminous and dark sectors, at current  experimental level of sensitivity.  We consider several different DM  phenomenological models from the myriad of those presently allowed. In this contribution we review astrophysical aspects of interest in the interplay of ordinary matter and a fermionic light Dark Matter component. We focus in the interior nuclear medium in the core and external layers, i.e. the crust, discussing the impact of a novel dark sector in relevant stellar quantities for (heat) energy transport such as thermal conductivity or emissivities.
\end{abstract}

\end{frontmatter}

\section{INTRODUCTION }
\label{sec:intro}
As early as 1933 the discovery made by F. Zwicky about visible matter being only a tiny fraction of the matter in our Universe, evidenced that our understanding of the cosmos was far from being complete. Since that date a variety of further indications including galactic rotation curves, anisotropies in the cosmic microwave background, distribution of galaxies, etc, points to the existence of cold non-luminous sector of particles, i.e. Dark Matter (DM), see \citet{evidences1,vera, bullet}. Nowadays, this paradigm is well accepted by the scientific community and we know DM  constitutes the most abundant type of matter in our Universe, being its density experimentally well-determined, $\Omega_{CDM}h^2= 0.1199 \pm 0.0027$ by the \cite{cdm}, although its true identity remains unknown, see for example a discussion in \cite{bertone}. 

The Standard Model (SM) of particle physics alone is not capable to explain the nature of this DM, suggesting that it must be extended. Current  experimental constraints come from attempting different strategies. Direct detection searches of thermalized galactic DM ($\chi$) are mainly based on nuclear recoils on selected targets. For the interaction cross-sections with nuclear matter (i.e. nucleons, $N$), there are at least five orders of magnitude remaining to be fully tested, namely $\sigma_{\chi N}\sim 10^{-43}$-$10^{-48}$ cm$^2$  in the $m_\chi \sim$ few GeV mass scale. DM particle existence could also be deduced from their annihilation products in indirect search experiments. In the case of a Majorana candidate, the indirect signal expected at Earth involving detection of gamma-rays  or neutrino final products is still under debate, see \cite{indirec}. A possible (tiny)  modification of standard physics in systems of interest is another way to indirectly detect its presence. Along this direction rely other lines of research such as the modification of the emissivity of Standard Model neutrinos (or steriles) in solar reaction chains that have been suggested by \cite{lopes-silk}. In addition, colliders aim to produce DM candidates through decay or annihilation of SM particles \citet{limit1, limit2}. 

In the light (low) mass region of DM candidates (LDM), $m_{\chi}\lesssim 1\; \rm GeV/c^2$, cosmological, astrophysical and collider constraints  seem to be the most significant, due to the fact that particles with masses smaller than that of the nucleon can only provide kinematical recoil energies a the $\sim$ eV range, below the $\sim$keV threshold for current conventional terrestrial searches in direct detection experiments, see \cite{lin}. If one, instead, considers LDM scattering off bound electrons, energy transfer can cause excitation or even ionization and thus seems promising for exploring  the phase space in a complementary way, as explained in \cite{ele}. Contrary to what has been done in cosmological analysis, the importance of the dark sector component  on celestial body scales has been less extensively studied and only recently has  focused on the sun  \citet{sun}, planets or to a lesser extent on white dwarfs (\citet{wddm, wddm2}) and compact stars (\citet{nsdm, nsdm2}).
\section{CAPTURE OF DARK MATTER BY NEUTRON STARS}
DM reaching terrestrial targets is expected to have low velocities  $\beta_{\rm Earth}=\sqrt{\frac{2GM_{\rm Earth}}{R_{\rm Earth}}}$ = $v_\chi \sim 10^{-3}$ (taking $c=\hbar=1$ units and being $G$ the gravitational constant) due to the fact that the gravitational boost is small for the Earth and, accordingly, the Lorentz factor $\gamma_{Earth} =1/\sqrt{1-\beta_{Earth}^2} \sim 1$. Instead, for more compact objects i.e. with a larger ratio $M/R$, the boosting capability increases. For neutron stars (NSs) with masses $M_{\rm NS}\simeq 1.5 M_\odot$ and radius  $R_{\rm NS}\simeq 12$ km, $\gamma_{\rm NS}\sim 1.26$ or $v_\chi \sim 0.6$, providing   thus a mechanism to attain higher $\chi$ kinetic energies.

It is important to remark that DM could be in principle accreted (and retained) by the star not only during the collapsed stage but also during most of the previous stellar lifetime, being the capture rate, $C_{\chi}$, provided the interaction strength with ordinary matter is not negligible. Different particle candidates display interaction couplings in a wide range \cite{bertone}. Regarding the progenitor stages, the progressively denser central core allows the rise of a finite spatial  number density, $n_\chi(r)$, over time. Later, when a compact object is formed the efficiency changes and the capture rate enhances, see for example \cite{gould, kouv}. In this way, assumed an equation of state for regular SM matter in the interior of the NS, at a given galactic location, and with a corresponding ambient DM mass density $\rho_\chi \sim m_\chi n_\chi$, following \citet{nsdm2, guver}, the capture rate can be written for a weakly interacting DM particle as
\begin{eqnarray}
C_{\chi}&=&\frac{8}{3}\pi^2 \frac{\rho_{\chi}}{m_{\chi}}\frac{GM_{NS}R_{NS}}{1-\frac{2GM_{NS}}{R_{NS}}} \bar{v}^2 \left( \frac{3}{2\pi \bar{v}^2} \right)^{\frac{3}{2}} f_{\chi} \nonumber \\  & \simeq & 1.8 \times 10^{25} \left(\frac{1\, \rm GeV}{m_{\chi}}\right)\left(\frac{\rho_{\chi}}{\rho_{\chi,0}}\right) f_{\chi}\,\,\rm s^{-1},
\label{capt}
\end{eqnarray}
%
\normalsize
where $\bar{v}$ is the average $\chi$ velocity in the existing distribution, with a local value $\rho_{\chi,0}\simeq 0.3$ $\rm \frac{GeV}{cm^3}$ and $f_\chi$ is the fraction of particles that undergo one or more scatterings while inside the star. It saturates to $f_\chi \sim 1$ if ${\sigma_{\chi N}}\gtrsim {\sigma_0}$, otherwise $f_\chi \sim 0.45 \frac{\sigma_{\chi N}}{\sigma_0}$, being $\sigma_0=\frac{m_n R_{\rm NS}^2}{M_{\rm NS}}\sim 10^{-45}$ $\rm cm^2$ the geometrical cross-section. As thermalization times for $m_\chi\sim 1$ GeV particles are much smaller than dynamical cooling times, see \cite{nsdm}, once inside, DM is believed to diffuse toward the denser central stellar regions according to the exponential law 
\begin{equation}
n_{\chi}(r)=n_{0,\,\chi} e^{-\frac{m_{\chi}}{k_B T}\Phi (r)}
\end{equation}
 with $n_{0,\,\chi}$ the central value and $\Phi(r)=\int_0^r \frac{GM(r') dr'}{{r'}^2}$ the gravitational potential. Finally $n_{\chi}(r)=n_{0,\,\chi} e^{-(r/r_{\rm th})^2}$ with a thermal radius $r_{\rm th}= \sqrt{\frac{9 k_B T}{8 \pi G \rho_n m_{\chi}}}$, where $\rho_n$ is the barionic density. Normalization requires $\int_0^{R_{NS}} n_{\chi}(r) dV=N_\chi$ at a given time. $N_\chi$ is the number of DM particle population which resides inside. This number will depend, in general, on the capture, annihilation (or decay) and evaporation rates $C_\chi$, $C_a$, $C_s$, respectively, see for example \cite{zentner}. But, for DM particles with masses larger than $\sim 2 \; \rm keV$ evaporation effects can be ignored as it is shown in \citet{kouv&tin, evap}. Then, as a function of time $t$, capture and annihilation processes compete to yield a population 
\begin{equation}
N_\chi(t)=\sqrt{C_\chi/C_a} \rm tanh (t/\tau+\gamma (N_{\chi,0}))
\end{equation}
 where $\gamma (N_{\chi,0})=\mathrm {tanh^{-1}} (\sqrt{C_a/C_\chi}N_{\chi,0}) $ and $\tau=1/\sqrt{C_\chi C_a}$. For a typical progenitor of a compact star, one can estimate $ N_{\chi,0} \sim 10^{39} (\frac{m_\chi}{1\, \rm GeV})$, see \cite{kouv&tin2}, assuming that local densities for average NS galactic distances peak around $\sim 2$ kpc where $\rho_{\chi}\sim 10^2 \rho_{\chi,0}$. For times $t \gtrsim \tau \sim 10^{4}$ yr the equilibrium sets $N_\chi(t)\sim \sqrt{C_\chi/C_a}$.

In the case of asymmetric DM candidates, accretion of DM mass beyond a critical value, i.e. the Chandrasekhar mass, could destroy (collapse) the star over time as it has been studied in \citet{kouv, zurek, bramante}.  Possible limiting values of $N_\chi$ arising from a fermionic nature provide a value of  Chandrasekhar number  $N_{\rm Ch}\sim (\,M_{\rm Pl}/m_{\chi})^3\sim 1.8 \times 10^{54}\,(\rm1 \,TeV/m_{\chi})^3$ with $M_{\rm Pl}$ the Planck mass.

Another way quoted to limit the dark sector population and thus collapse the whole star stems from the fact that quark deconfinement could be triggered via DM seeding. Energy release from self-annihilating DM could induce spark formation energetic enough to  nucleate stable bubbles of deconfined quark matter leading to a softening of the nucleon equation of state. This has been invoked as a mechanism that could drive NS to quark star (QS) conversion, see \citet{perez-silk1, perez-silk2, perez-silk3}. In addition, arguments along the same line state that  unstable DM can also be constrained by structural stability of  accreting objects \cite{perez4}.

\section{LIGHT DARK MATTER SCATTERING  AT HIGH DENSITIES}

The DM capture rate Eq. \eqref{capt} mainly determines the magnitude of the novel effects due to the interplay of both matter sectors.  For a given DM candidate, the interaction with the relevant degrees of freedom of the system considered will be crucial. This, in turn, comes phenomenologically described through couplings and mathematical operators in the underlaying theory. So far no satisfactory answers exist coming from a unique model description, although there is a general consensus that the particle or particles in the dark sector should be electrically neutral and cold.  The most popular extension to the Standard Model, supersymmetry (SUSY), predicts that each particle has a partner of different spin but similar interactions. The lightest superpartner or LSP is stable in many cases and is often a neutralino, constituting an excellent dark matter candidate \cite{bertone}.

In a nuclear medium such as the core of a compact-sized (spherical) object of mass $M$ and radius $R$,  the (in) elastic scattering of $\chi$ out of nucleons,  $\sigma_{\chi N}$, mainly determines the relative fraction of DM to ordinary matter.  In general, most of the works calculate this quantity without considering  {\it in-medium} effects, and obtain the mean free path of these DM particles, $\lambda_\chi$, approximating as $\lambda_\chi\simeq 1/\sigma_{\chi N} n$ where  $n$ is the ordinary nucleon number density, see for example \cite{spergel}. This is usually considered as being  sufficient to obtain knowledge about the internal processed happening inside the most efficient opaque environments. It is also important to know that the isospin charge is relevant as the coupling strengths to protons  or neutrons may not be the same or even suffer from cancellations. 

In the central core in a NS (with a content $\gtrsim 90\%$ neutrons), is described as having densities higher than several times nuclear saturation density $n_0\simeq 0.17$ $\rm fm^{-3}$. By neglecting medium corrections one misses important features and qualitative insight into the system. Let us briefly comment on some of the possible issues. To begin with, Fermi-blocking, due to partial restriction of the outgoing nucleon phase space, can play a role diminishing the ${\chi N}$ scattering cross-section. Finite temperature effects will  allow the population of higher energy states in the nucleon sector with respect to the case in which no temperature effects are considered. It is important to note here that temperatures in the range $T\lesssim 50$ MeV are usually achieved in the very early stages of proto-neutron star evolution, see for example \cite{page04}. Later, after a primary neutrino cooling era, temperatures fall to the $\sim$ keV range. This will effectively set at large times a $T\approx0$ configuration, as thermal energies are indeed much smaller than nucleon Fermi energies  $k_{\rm B}T<<E_{\rm FN}$ in the dense medium ($k_{\rm B}$ is the Boltzman constant). These in-medium effects are taken into account, for example, in \citet{bertoni, cermeno1, cermeno2}.

As we have mentioned before, usually, the incoming DM particle is supposed to be thermalized in the galaxy with the Maxwellian mean velocities $v_\chi \sim 10^{-3}$. However, if the scenario considered is an accreting dense NS  the associated  wavelength of the incoming DM particle decreases as $\lambda=\frac{2 \pi}{\sqrt{\gamma^2 -1}\,m_\chi }$. This expression sets, in practice, a measure of the validity of the quasielastic approximation, DM particles see nucleons as a whole, since matter is tested to sizes around $\lambda\sim$ 1 fm, i.e. in the light DM mass range $m_\chi\lesssim 5$ GeV. Although further modeling would be required for the description of the inner hadron structure, for this mass range, a monopolar form factor in momentum space, $F(|{\vec q}|)=\frac{\Lambda^2}{\Lambda^2+q^2}$ with a cut-off parameter $\Lambda=1.5$ GeV can be used, so that $F(0)=1$, which can somewhat mitigate the short-range correlations arising in the calculation.

%

To check the importance of the modifications of the medium in the LDM scattering events one has to calculate the differential cross-section per unit volume for the interaction between a fermionic particle of Dirac type, $\chi$, and a nucleon field $N$ (protons and neutrons). It is necessary to use  the general expression, which can be found in \cite{pdg},
\begin{equation}
d\sigma=\frac{|\mathcal{\overline{M}_{\chi \rm N}}|^2|F(|{\vec q}|)|^2}{4\sqrt{(p_N p_\chi)^2-{m^*}_N^2 m_{\chi}^2}}d\Phi(p_N,p'_N,p_\chi,p'_\chi)\mathcal{F_{FB}},
\label{dsigma}
\end{equation}
where the phase space volume element is
\begin{equation}
\begin{aligned}
d\Phi(p_N,p'_N,p_\chi,p'_\chi)= & (2\pi)^4\delta^{(4)}(p_N+p_\chi-p'_N-p'_\chi)\\ 
& \times \frac{d^3\vec{p'_N}}{(2\pi)^32E'_N}\frac{d^3\vec{p'_\chi}}{(2\pi)^32E'_\chi}.
\end{aligned}
\end{equation}
$p'^{\mu}_N=(E'_N, \vec{p'_N})$ and $p^{\mu}_N=(E_N, \vec{p_N})$ are the four momenta for the outgoing and incoming nucleon, and $p'^{\mu}_\chi=(E'_\chi, \vec{p'_\chi})$ and $p^{\mu}_\chi=(E_\chi, \vec{p_\chi})$ those analogous for the DM particle, respectively. Momentum transfer is denoted by $q^{\mu}=p'^{\mu}_\chi-p^{\mu}_\chi$. In this way $q_{0}=E'_N-E_N=E_\chi-E'_\chi$ and $\vec{q}=\vec{p'_N}-\vec{p_N}=\vec{p_\chi}-\vec{p'_\chi}$. 
The four-dimensional delta assures the energy and momentum conservation in the collision and $|\mathcal{\overline{M}_{\chi \rm N}}|^2$ is the squared matrix element for the interaction considered. The factor $\mathcal{F_{FB}}=f(E_N)(1-f(E'_N))$, with $f(E_i)=\frac{1}{1+e^{({E_i-\mu^*_i})/{k_{B}T}}}$ $i=$p,n and $\mu^*_i$ the effective nucleon chemical potential for a particle with isospin of $i$th-type, see \cite{reddy}, accounts for the Fermi blocking term. It takes into account the occupation of states and  in this scenario affecting  only to the nucleons. For the dark sector, it can be assumed that all outgoing DM particles states are in principle  allowed and $1-f_\chi(E)\approx 1$ since the fraction of DM inside the star remains tiny at all times. The validity of this approximation is discussed in \cite{cermeno1}.  
Let us mention here that effective values of nucleon mass and chemical potential define the {\it quasi-particle} nature of the nucleon in the medium and differ from the nude values by the presence of average meson fields \cite{wal}. 

When discussing the effect of density (and $T$) in the diffusion of DM in NSs, the quantity we should analyse is the mean free path $\lambda_\chi$ to be obtained from the integrated cross-section per unit volume  $\frac{\sigma(E_\chi)}{V}$, as $\lambda_\chi=\left(\frac{\sigma(E_\chi)}{V}\right)^{-1}$. 

In a previous step one obtains the differential value as, 
\begin{equation}
\begin{aligned}
\frac{1}{V}\frac{d\sigma}{d\Omega dq_0}=&\frac{1}{(2\pi)^4} \int_{|\vec{p}_-|}^{\infty}\frac{d|\vec{p_N}||\vec{p_N}|}{4E'_N}\frac{m^*_N|\vec{p'_\chi}||F(|{\vec q}|)|^2}{|\vec{q}|}\\ 
&\times \delta(cos\; \theta-cos\; \theta_0)\Theta(|\vec{p_N}|^2-|\vec{p_N}_-|^2) \\
&\times \frac{|\mathcal{\overline{M}_{\chi N}}|^2 f(E_N)(1-f(E'_N))}{4\sqrt{E_N^2E_\chi^2-{m^*}_N^2m_{\chi}^2}} ,
\end{aligned}
\label{difsigma}
\end{equation}
where only the lowest order terms are retained.  Since $v^2_{\chi}\sim v^2_N\ll1$ given $\frac{|\vec{p_i}|}{E_i}=v_i$, $i=\chi, N$ and nucleon Fermi velocities $v_N\sim v_{FN}=p_{FN}/E_{FN}$.
On the other hand, expressing the energy conservation as a function of the dispersion angle for the outgoing DM particle $\theta$, one obtains the minimun value of $\vec{p_N}$
\begin{equation}
cos\; \theta_{0}=\frac{m^*_{N}}{|\vec{p}||\vec{q}|}\left( q_{0}-\frac{|\vec{q}|^2}{2m^*_N}\right) ,
\label{cos}
\end{equation}
and 
\begin{equation}
|\vec{p}_-|^2=\frac{{m^*_N}^2}{|\vec{q}|^2}\left( q_{0}-\frac{|\vec{q}|^2}{2m^*_N}\right)^2.
\label{pn}
\end{equation}
These expressions, Eqs. (\ref{cos}) and (\ref{pn}), are discussed step by step in \cite{cermeno1}.\\
We have to remark that this scenario is restricted to temperatures and densities typical for the  thermodynamical evolution of the stellar core region, that is $T \lesssim 50$ MeV and $n\simeq (1-3)n_0$. Let us note that if finite temperature is considered, a detailed balance factors must be added  to the medium response to weak probes (\cite{paper6, horo}) under the form
\begin{equation}
S(q_0,T)=\frac{1}{1-e^{-\frac{|q_0|}{k_{B}T}}} .
\label{fbalance}
\end{equation}
This factor provides the relation between the dynamical nuclear structure factor for positive and negative energy transfers $q_0$ as the thermodynamic environment can donate energy to the outgoing particle. 

Now, to obtain the total cross-section per unit volume and the inverse of it, i.e. the  mean free path, the integrations over all possible outgoing energy transfer values and solid angle have to be performed. In this way,
\begin{equation}
\begin{aligned}
&  \frac{\sigma(E_\chi)}{V}=\frac{m^*_N}{4(2\pi)^3}\int_{0}^{E_\chi-m_{\chi}} dq_0  \int_{|\vec{p_\chi}|-|\vec{p'_\chi}|}^{|\vec{p_\chi}|+|\vec{p'_\chi}|} d|\vec{q}||F(|{\vec q}|)|^2 \\
&\times \int_{|\vec{p}_-|}^{\infty}d|\vec{p_N}|\frac{|\mathcal{\overline{M}_{\chi N}}|^2|\vec{p_N}|f(E_N)(1-f(E'_N))S(q_0,T)}{4E'_N|\vec{p_\chi}|\sqrt{E_N^2E_\chi^2-{m^*_N}^2m_{\chi}^2}}.
\label{sigma}
\end{aligned}
\end{equation}

When particularizing the calculation for a spin-independent (scalar-vector) interaction model the interaction Lagrangian is $\mathcal{L_I}=\sum_{N=n,p}g_{s,N}\chi\overline{\chi} N \overline{N}+g_{v,N}\chi\gamma^{\mu}\overline{\chi} N\gamma_{\mu} \overline{N}$ where $g_{s,N}$ $(g_{v,N})$ are the effective scalar (vector) couplings of the DM particle to the $N$ field.

\begin{figure}[ht]
\begin{center}
\includegraphics [angle=0,scale=1.3] {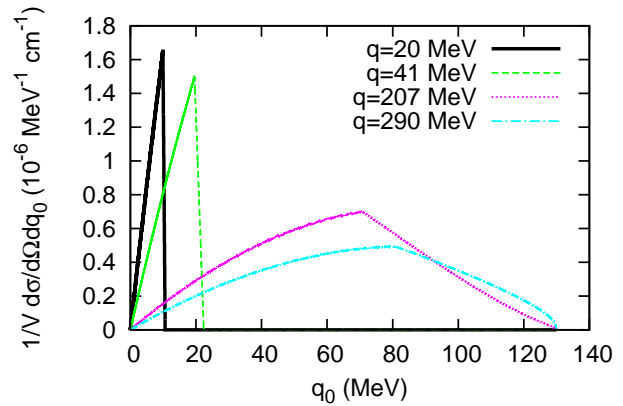}
\caption{Differential cross-section per unit volume as a function of the energy transfer $q_0$ for values of $|\vec q|=20, 41, 207$ and $290$ MeV. The DM particle mass is $m_\chi=0.5$ GeV and $T=0$ at $n=n_{0}$. From \cite{cermeno1}.}
\label{fig1dif}
\end{center}
\end{figure}

The effect of the Pauli blocking can be seen, for example, in  Fig. (\ref{fig1dif})  of  \cite{cermeno1} where the differential cross-section per unit volume as a function of the energy transfer $q_0$ for different values of the transferred momentum $|\vec{q}|$ is shown for a pure neutron system. We use $n=n_{0}$, setting $T=0$ and $m_\chi=0.5 \; \rm GeV$. The limiting upper value of the energy transfer is $E_\chi-m_\chi \approx 130$ MeV. 
This is because, in general, $-\infty<q_0<E_\chi-m_{\chi}$, since $m_{\chi}<E'_\chi<\infty$. Instead, at  $T=0$, the energy transfer can not be negative, so that $q_0>0$ and  $0<q_0<E_\chi-m_{\chi}$. 
The triangular shape is due to the Heaviside Fermi distribution at $T=0$. Beyond $q_0$ values limited by real-valued angles in Eq.(\ref{cos}) the scattered states are not allowed since it is kinematically impossible to scatter a nucleon due to lack of empty states.
To understand the value of the maxima of this plot we have to remind ourselves that at $T=0$ our allowed states will be given by the factor $\Theta(E_F-E_N)\Theta(1-f_N(E_F-E'_N))$, and, on the other hand, there is a minimun allowed value for $\vec{p_N}$ given by the kinematics of the interaction, $|\vec{p}_{-}|$. Then, it is easy to verify that maximum values are reached when $|\vec{p}_{-}|=|\tilde{p_N}|$ being $|\tilde{p_N}|$ the value of $|\vec{p_N}|$ when $|\vec{p'}_N|=|\vec{p}_F|$. 

\begin{figure}[ht]
\begin{center}
\includegraphics [angle=0,scale=1.3] {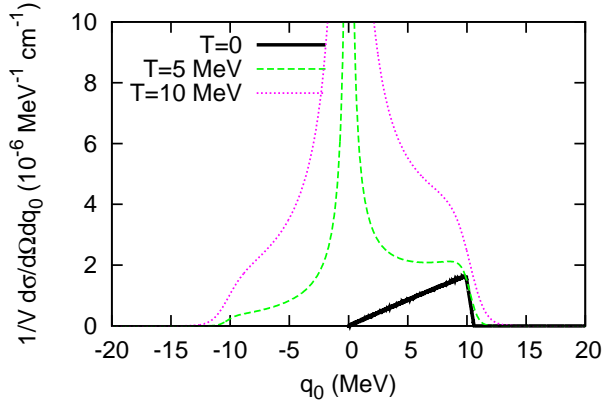}
\caption{Differential cross-section per unit volume as a function of the energy transfer $q_0$ at $T=0, \,5,\, 10$ MeV for a nucleon density $n=n_{0}$. We set  $|\vec q|=20$ MeV and $m_\chi=0.5$ GeV. From \cite{cermeno1}.}
\label{fig4}
\end{center}
\end{figure}

\begin{figure}[ht]
\begin{center}
\includegraphics [angle=0,scale=1.3] {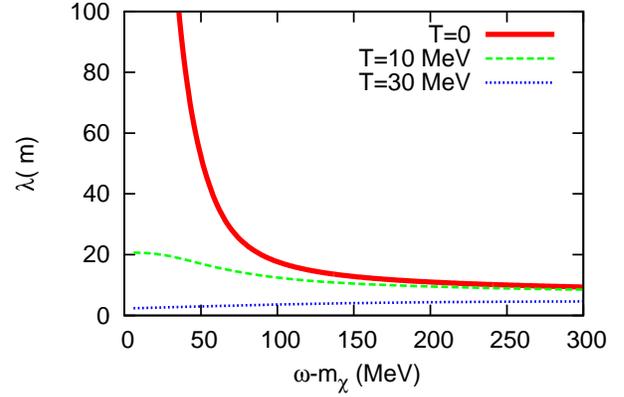}
\caption{DM particle mean free path as a function of kinetic energy for $m_{\chi}=1$ GeV at $n=n_{0}$ for $T=0, 10, 30$ MeV. From \cite{cermeno1}. }
\label{fig8}
\end{center}
\end{figure}

Finite temperature effects can be observed in  Fig. (\ref{fig4}) where the detailed balance factors have been included and the transferred momentum $|\vec q|=20$ MeV, chemical potential $\mu=E_{FN}$, DM particle mass $m_\chi=0.5 \; \rm GeV$ and nucleon density $n=n_0$ are fixed.\\ 
At temperatures $T>0$ the negative energy transfer states get increasingly populated and the sharp nucleon distribution is smoothed. As $q_0 \rightarrow 0$ the inverse detailed balance factor $S^{-1}(q_0, T) \rightarrow 0$. The corresponding divergence will, however, be integrable in order to obtain a finite integrated cross-section.
This, can be seen, for example, in Fig. (\ref{fig8}), where the mean free path for the $\chi$ particle as a function of kinetic energy $K_\chi=E_\chi-m_{\chi}$ for temperatures $T=0$ (solid line) $T=10$ MeV (dashed line) and $T=30$ MeV (dotted line) is plotted. $m_{\chi}=1$ GeV and $n=n_{0}$ are considered. The notation used states that $E_\chi\equiv \omega$. Note that for $T=0$ in the limit of vanishing kinetic energy for the incoming DM particle $E_\chi \rightarrow m_{\chi}$, the phase space available for the outgoing particles vanishes as the energy transferred $q_0 \rightarrow 0$ due to filled population levels, therefore providing a null cross-section, while this is not true in the finite $T$ case as more channels are available.

\begin{figure}[ht]
\begin{center}
\includegraphics [angle=0,scale=1.3] {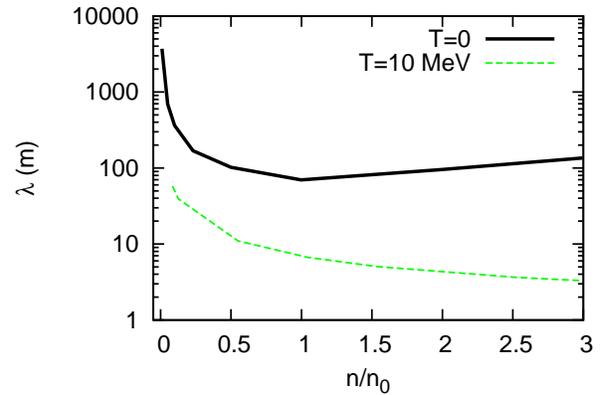}
\caption{DM particle mean free path as a function of density (in units of $n_{0}$) for two values of temperature, T=0 and T=10 MeV. Effective (naked) nucleon mass has been used in the zero (finite) T calculation. From \cite{cermeno1}.}
\label{fig7}
\end{center}
\end{figure}

In Fig. (\ref{fig7}), we show  the variation of the DM particle mean free path as a function of density (in units of $n_{0}$) for two values of temperature and fixing $m_\chi=1$ GeV considering effective nucleon masses for the $T=0$ case while not for the finite temperature case. We can see competitive effects from the temperature and effective mass. A steady decrease is obtained in case the naked nucleon mass is considered. Temperature effects, which are relevant in the early stages of dense star evolution,  tend to increase the opacity of nucleon matter to prevent DM nearly-free streaming. The cross-section is greatly affected by the finite density of matter, namely by the effect of a smaller  effective nucleon mass $m^*_N<m_N$. Temperature effects are important  although to a lesser extent relative to density ones.


It is important to point out that in this LDM mass range, scattering is diffusive to very good approximation as $\lambda_\chi/R_{\rm NS} \ll 1$. On the other hand, for the choice of couplings strengths consistent with collider constraints as \citet{limit1, limit2}, once fixed $K$ energy, a Standard Model neutrino displays typical mean free path somewhat smaller, see \cite{reddy,horo,paper6}.
\section{LIGHT DARK MATTER SCATTERING  AT LOW DENSITIES}

At low densities $\lesssim 10^{13}$ $\rm g/cm^3$ inside the star crust matter arranges in a series of irregular shapes or structures called generically {\it pasta phases} \cite{pasta}. In particular,  the outer crusts in NSs are formed by periodically arranged nuclei with typical densities ranging from $\rho \simeq  2 \times 10^6-4 \times 10^{11}$ $\rm g/cm^3$. In the single-nucleus description, see \cite{ruster}, a series of nuclei with increasing baryonic number, $A$, from Fe to Kr form a lattice before neutrons start to leak out of nuclei. At these high densities, electrons form a degenerate Fermi sea. Around $\rho_0\simeq n_0 m_N \simeq 2.4\times 10^{14}$ $\rm g/cm^3$, and beyond that, matter can be considered a homogeneous system.

When the phase space is restricted to fermionic DM particles with masses $m_\chi \lesssim 100 \;\rm MeV$, the relevant cross-section of the interaction between DM particles and the outer stellar layers is that scattering off nuclei. 
Taking this into account, the possibility of production of quantized lattice vibrations (phonons) in the periodically arranged structures in the NS outer crust by LDM scattering, arises. These extra phonons can impact subsequent quantities of interest, such as the ion thermal conductivity, that are relevant for computing the cooling behavior of NSs.
 
Phonons are quantized vibrational modes characterized by a momentum $\vec{k}$ and polarization vector $\vec{\epsilon}_{\lambda}$ appearing in a nuclear periodic system, as it can be seen in \cite{ziman}. They can have a number of different sources. They can be excited due to non-zero temperature $T$ in the medium. The Debye temperature allows us to evaluate the importance of the ion motion quantization. For a bcc lattice, see \cite{carr}, $T_D\simeq 0.45 T_p$, being $T_p=\omega_p /k_{\rm B}=\sqrt{\frac{4 \pi n_AZ^2e^2}{k_{\rm B}^2 m_A}}$ the plasma temperature associated to a medium of ions with number density $n_A$, baryonic number $A$, electric charge $Ze$ and mass $m_A$. 
At low temperatures $T<T_D$, the quantization becomes increasingly important and the thermal phonons produced are typically acoustic modes, following a linear dispersion relation $\omega_{{k},\lambda}=c_{l, \lambda}|{\vec k}|$, where $c_l=\frac{\omega_p/3}{ (6\pi^2 n_A)^{1/3}}$ is the sound speed. 
In addition, phonon production can be caused by an external scattering agent, for example, standard model neutrinos. In this respect, weak probes such as cosmological neutrinos with densities $n_\nu \sim 116$ $\rm cm^{-3}$ per flavor have been shown to provide small phonon production rates in a crystal target, see \cite{ferreras}. Due to the tiny mass of the neutrino, the experimental signature of this effect seems however hard to confirm.

The single phonon excitation rate (per unit volume) is obtained as \cite{cermeno2}
\begin{equation}
R^{(0)}_{{k}}=\frac{8\pi^4n^2_A }{(2 \pi)^6 m_{\chi}^2m_Ac_l} \int_{0}^{\infty} |\vec{p_\chi}|d|\vec{p_\chi}| f_{\chi}(\vec{p_\chi})|E_\chi-|\vec{k}|c_l|a^2,
\end{equation}
where $f_{\chi}(\vec{p_\chi})=\frac{n_\chi \mu }{4 \pi m_\chi^3 K_2(\mu)} e^{-\mu \sqrt{1+\frac{|\vec{p_\chi}|^2}{m_\chi^2}}}$ is the Maxwell-Juttner distribution (\cite{juttner}) function for relativistic incoming DM particles. With $\mu=\frac{m_\chi}{k_{B\rm}T}\approx 6.7$ for $\sqrt{<v^2>}\sim 0.6$, see \cite{Hakim} and \cite{Cercignani}, and $K_2(\mu)$ the modified Bessel function of second kind. 
The value of $a$ takes into account the length of the interaction when a DM particle approaches a nucleus in the periodic lattice. Using the Born approximation for the cross-section (off protons and neutrons) in the center of mass frame, it can be written as
\begin{equation}
4 \pi a^2= {m^2_A} \frac{\left(\frac{ Z}{m_p}\sqrt{|\tilde{\mathcal{M}_p}|^2}+\frac{(A-Z)}{m_n}\sqrt{|\tilde{\mathcal{M}_n}|^2}\right) ^2}{16 \pi (m_\chi+ m_A)^2},
\label{se}
\end{equation}
with  $\int_{-1}^{1} 2\pi d( cos \; \theta_{\chi}) |\mathcal{\overline{M}_{\chi \rm N}}|^2 \equiv  |\tilde{\mathcal{{M}_{\rm N}}}|$ and the averaged matrix element considering a prescribed model interaction. In this expresion, Pauli blocking effects are not considered but, as an approximation, density dependence can be retained using a parametrization of the nuclear Fermi momentum $|\vec{p}_{\rm FN}| \sim(3 \pi^2 n_0 Y_N)^{1/3}$ and the nuclear fractions $Y_p=Z/A$, $Y_n=(A-Z)/A$. 
It is important to say that the allowed values for the phonon momentum are restricted for the Born approximation, $|\vec{k}|\ll \frac{1}{a}$, as well as for kinematical restrictions when energy conservation is imposed, $0\leq |\vec{k}|\leq 2 \left(  \frac{c_lE_\chi }{(c_l^2-1)}+\frac{|\vec{p_\chi}|}{|c_l^2-1|}\right)$. 

The single phonon excitation rate (per unit volume) $R^{(0)}_{{k}}$ due to the accretion of DM particles is almost constant with $|\vec{k}|$ whereas for neutrinos $ R^0_\nu(|\vec{k}|)=R_{\nu 0} e^{  -b|\vec{k}|}$ with $b$ a constant which depends on neutrino mass. This means that, for a fixed value of $|\vec{k}|$ contribution from DM particles to the phonon excitation rate will be much higher than neutrino contribution.

Now, it will be important to discuss the astrophysical impact of these phonons. As we have mentioned before, phonon production  can be crucial for determining further transport properties, in particular, thermal  conductivity in an ion-electron system such as that in the outer NS crust. As an important contribution to the total ion conductivity, $\kappa_i$, partial ion conductivities due to ion-ion, $\kappa_{ii}\equiv \kappa_{ph}$, and ion-electron collisions, $\kappa_{ie}$, must be added, see for example \cite{chugunov}, under the prescription  $\kappa^{-1}_i=\kappa^{-1}_{ii}+\kappa^{-1}_{ie}$. Standard mechanisms to produce lattice vibrations include thermal excitations, as analyzed in detail in previous works \citet{negele, pot, baiko}. In a NS, the outer crust  can be modeled under the one-component-plasma description. This low density solid phase can be classified according to the Coulomb coupling parameter $\Gamma=Z^2e^2/a k_{\rm B} T$ where $a=(4 \pi n_{A}/3)^{1/3}$ is the ion sphere radius. It is already known that typically for $\Gamma \ge \Gamma_m\simeq 175$, or below melting temperature $T<T_m$, single-ion systems crystallize, see \cite{gamma}.

There are a number of processes that can affect thermal conductivity in the medium. The so-called U-processes are responsible for  modifying the electron conductivity such that for high temperatures, $T>T_U$,  electrons move almost freely, more details in \cite{ziman}. Assuming a bcc lattice, $T_U\simeq 0.07T_D$. Thus in the scenario depicted here,  the  temperature range must be $T_U <T<T_D<T_m$ for each density considered.  According to kinetic theory, the thermal conductivity can be written in the form 
\begin{equation}
\kappa_{ii}=\frac{1}{3} k_{\rm B} C_A n_A c_l L_{ph},
\end{equation} 
as it can be seen in \cite{ziman}, where $C_A=9 \left(\frac{T}{T_D}\right)^3 \int_0^{T_D/T} \frac{x^4 e^x dx}{(e^x -1)^2}$ is the phonon (dimensionless) heat capacity (per ion) and $L_{ph}$ is an effective  phonon mean free path that includes all scattering processes considered:  U-processes and impurity (I) scattering processes (both dissipative) and the phonon normal (N) scattering which are non dissipative $L^{-1}_{ph}=L^{-1}_U+L^{-1}_I+L^{-1}_N$. Typically the thermal conductivity is related to the thermal phonon number  at temperature $T$, $L_{ph}\sim 1/N_{0,k \lambda}$ where $N_{0,k \lambda}=(e^{\omega_{k\lambda}/k_{\rm B} T}-1)^{-1}$. The contribution from DM can be obtained by the net number of phonons that results from the competition between thermal phonons and scattering excitation and stimulated emission, as it is shown in \cite{ferreras}, in a 4-volume $\delta V \delta t$. Using the averaged rate per unit volume and weighting with the incoming distribution providing the frequencies of different values of momenta the total number of phonons can be written as 

\begin{equation}
\begin{aligned}
N_{k\lambda} & \simeq  N_{0,k\lambda}+ R^{(0)}_k \delta V \delta t \\
& - \int \frac{d^3\vec{p}}{n_\chi} \,f_{\chi}(\vec{p}) {\tilde R^{(0)}_k} N_{0,k\lambda} e^{(\omega_{k,\lambda}+ {\vec k}.{\vec v})/ K_\chi} \delta V \delta t,
\end{aligned}
\end{equation}
see \cite{cermeno2}. Where ${\tilde R^{(0)}_k}$ is the single phonon excitation rate for each particular momentum value (not averaged over incoming $\chi$ momenta). Since the source (NS) is in relative motion to the LDM flux, there is a Doppler shift characterized by the source velocity ${v}\equiv v_{NS}\sim 10^{-2}$ i.e galactic NS drift velocity. As we have mentioned before, the distribution of NSs in our galaxy peak at distances $\lesssim$ few kpc (\cite{lorimer}) where the DM density is enhanced with respect to the solar neighborhood value.

\begin{figure}[ht]
\centering
\includegraphics [angle=0,scale=1.3]{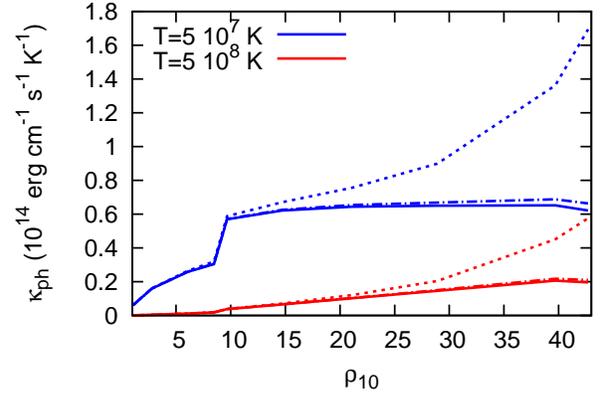}
\caption{Phonon thermal conductivity as a function of density (in units of $10^{10}$ $\rm g/cm^3$) for temperatures $T=5\,10^7$ K (blue), $5\,10^8$ K (red) and $m_\chi=100$ MeV. Dash-dotted and dashed lines depict the impact of a LDM density $n_\chi/n_{0\chi}=10, 100$. Solid lines are the standard thermal result with no DM for each case. From \cite{cermeno2}.}
\label{Fig2}
\end{figure}

In Fig. (\ref{Fig2}) the phonon thermal conductivity as a function of barionic density is shown for two different typical temperatures for the base of the crust, using $m_\chi=100$ MeV. Solid  lines are the standard thermal result with no DM. Dash-dotted and dashed lines correspond to $n_\chi/n_{0,\chi}=10, \,100$, respectively. 
It is clear to see that at the largest LDM local densities considered, there is an enhancement over the thermal result well inside the outer crust. This corresponds to the site where the DM-induced effects have the most influence (\cite{azorin}) as this is the most massive part of the outer crust. Although, below these densities, there is a negligible change. At lower $T$, the effect of a perturbation over the thermal phonon population is more important. Enhanced (decreased) conductivities at moderate LDM densities are due to a net reduction (increase) of the number of phonons in the lattice as a result of cancellation of modes. 

\begin{figure}[ht]
\centering
\includegraphics [angle=0,scale=1.3]{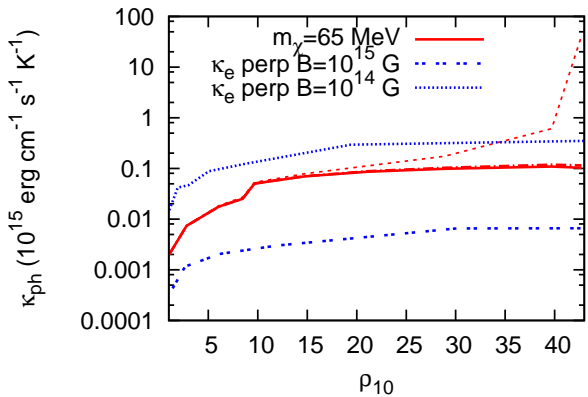}
\caption{Phonon thermal conductivity as a function of density (in units of $10^{10}$ $\rm g/cm^3$) at  $T=10^8$ K and $m_\chi=65$ MeV. Solid, dot-dashed and dashed lines correspond to cases with no DM, $n_\chi/n_{0,\chi}=10,\,100$. Perpendicular electron thermal conductivity is also shown for $B=10^{14}, 10^{15}$ G. From \cite{cermeno2}.}
\label{Fig3}
\end{figure}

The same type of plot can be shown in Fig.(\ref{Fig3}) in a more realistic stellar context.
Here, the phonon thermal conductivity as a function of baryonic density at  $T=10^8$ K and $m_\chi=65$ MeV can be seen. Solid, dot-dashed and dashed lines correspond to cases with no DM, $n_\chi/n_{0,\chi}=10,\,100$, respectively. Electron thermal conductivity is also shown for magnetized realistic scenarios  in the perpendicular direction to a magnetic field $B$ of strength $B=10^{14}$ G (dotted) and $B=10^{15}$ G (doble dotted), these NSs with so high magnetic fields are called magnetars and they are being studied by many groups, see for example a review from \cite{magn}. Ions are mostly not affected by the presence of a magnetic field for the range of temperatures and magnetic fields considered, this can be checked in \cite{chugunov}. The parallel direction electronic contribution is not depicted here since it is typically much larger $k_{e \parallel}\sim 10^{17}-10^{19}$ $\rm erg\,cm^{-1} \,s^{-1}\, K^{-1}$. Values of electronic contribution have been obtained from \cite{vigano}. Since the global conductivity is $\kappa=\kappa_e+\kappa_{ph}$, this result is expected to contribute to the reduction of the difference in heat conduction in both directions and thus to the isotropization of the NS surface temperature pattern, which is studied for example in \cite{aniso1, aniso2}, as it would be  smoothly driven towards more isothermal profiles for latitudes among pole and equator. It is already known, see for example \cite{kamin}, that the outer crust plays an important  role in regulating the relation among temperature in the base of it and the surface. The detailed calculation remains, however, for future works.
\section{IN-MEDIUM ENERGY RELEASE FROM ANNIHILATING DARK MATTER}

In previous sections we consider a scattering interaction between DM and ordinary matter. Apart from the spin-independent interaction model considered before (corresponding to effective scalar and vector couplings), more effective operators for fermionic DM particles can be studied. For example, the so-called Coy Dark Matter model proposed by \cite{Boehm} which belongs to the family of simplified models, see for instance ~\cite{Dolan2015, Bauer, Baek}, includes two new particles to the SM, i.e. a Dirac fermion DM candidate, $\chi$ which interacts with a boson $a$, with mass $m_a$, through the pseudoscalar coupling of strength $g_{\chi}$ as proposed in \cite{Arina}.
The interaction lagrangian of the model reads
\begin{equation}
\mathcal{L_I} =  - i \frac{g_{\chi}}{\sqrt{2}} a \bar{\chi} \gamma_5 \chi - i g_0 \frac{g_{f}}{\sqrt{2}} a \bar{f} \gamma_5 f .
\label{li}
 \end{equation}
A final state of fermion-antifermion is formed through the boson mediator. The vertex involves a pseudoscalar coupling  with the SM fermions $f$ with strength  $g_0g_{f}$, where $g_f$ denotes a multiplicative factor common to all couplings of $a$ with SM fermions. The most common type of couplings are the flavour-universal coupling $g_f=1$, independent of the fermion type, and Higgs-like coupling, proportional to the fermion mass $g_f=\frac{m_f}{174 \; \rm GeV}$. However, there are other schemes where $a$ couples either to quarks or leptons exclusively, and with a flavour structure \cite{Dolan2015}.

An existing DM distribution inside the star capable of self-annihilating can be a source of energy emission for the dynamical cooling behaviour, once the star is formed. In the range $m_\chi < m_{\rm Higgs}$ and $m_a < m_\chi$, the relevant annihilation processes are $\chi {\chi} \rightarrow f\bar{f}$ and $\chi {\chi} \rightarrow a a$, see \cite{Arina}.

In this context, these annihilation reactions, $\chi {\chi} \rightarrow f\bar{f}$ and $\chi {\chi} \rightarrow a a$, with subsequent decay $a \rightarrow f\bar{f}$ can be considered to explore possible astrophysical consequences in the neutrino fermionic channel as dense stars are efficient DM accretors. One of the key quantities that can govern the internal stellar energetic balance is the local energy emissivity, $Q_E=\frac{dE}{dVdt}$, which can be understood as the energy emitted per unit volume per unit time, through a prescribed particle physics reaction. 

Considering the above quoted reactions which involve annihilating DM accreted by a NS, with baryonic density $\rho_b\sim 2 \rho_0$, from an existing galactic distribution. $Q_E$ denotes the energy emission rate  per volume due to fermionic or pseudoscalar pair emission in the final states. In general, its expression can be written under the form (\cite{Esposito})
\begin{equation}
Q_E= 4\int d \Phi (E_1+E_2)\, |\overline{\mathcal{M}}|^2\, f(f_1,f_2,f_3,f_4),
\label{qe}
\end{equation}
with $d\Phi=\frac{d^3p_1}{2(2 \pi)^{3}E_1} \frac{d^3p_2}{2(2 \pi)^{3}E_2} 
\frac{d^3p_3}{2(2 \pi)^{3}E_3} \frac{d^3p_4}{2(2 \pi)^{3}E_4}\, (2 \pi)^4 \delta^4 (p_1+p_2-p_3-p_4)$ the 4-body ($12\rightarrow 34$) phase space element and $|\mathcal{\overline{M}}|^2 $ the spin-averaged squared matrix element of the reaction considered. The additional factor $f(f_1,f_2,f_3,f_4)$ accounts for the initial and final particle distribution functions contribution which will be different for each annihilation channel. The notation used is the following $p_1=(E_1, \vec{p_1})$, $p_2=(E_2, \vec{p_2})$ as the incoming 4-momenta while $p_3=(E_3, \vec{p_3})$, $p_4=(E_4, \vec{p_4})$ are the outgoing 4-momenta, respectively. Feynman diagrams and matrix elements can be seen in \cite{cermeno3}.

%
%
For the case of annihilation into fermionic pairs $f(f_1,f_2,f_3,f_4) =f_{\chi}(E_1)f_{\bar{\chi}}(E_2)(1-f_{f}(E_3))(1-f_{\bar{f}} (E_4))$  and $f_\chi,f_{f}$ are the local stellar distribution functions for DM and fermionic particles, respectively,  containing density and temperature dependence. Whereas for the annihilation into pseudoscalars the factor $f(f_1,f_2,f_3,f_4)=f_\chi(E_1)f_{\bar{\chi}}(E_2)f_a (E_3)f_a(E_4)$. 
\begin{figure*}[htp]
  \centering
  \subfigure{\includegraphics[scale=1.2]{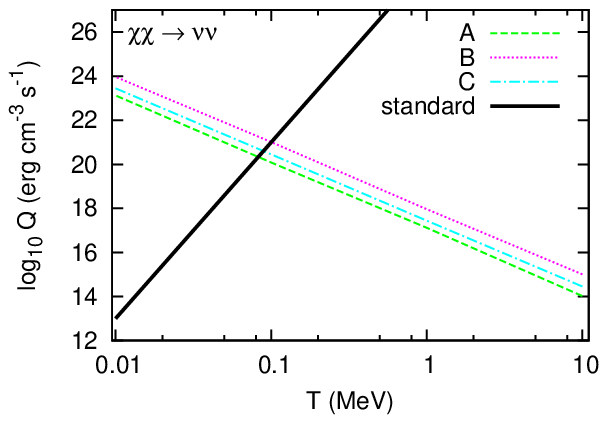}}\quad
  \subfigure{\includegraphics[scale=1.2]{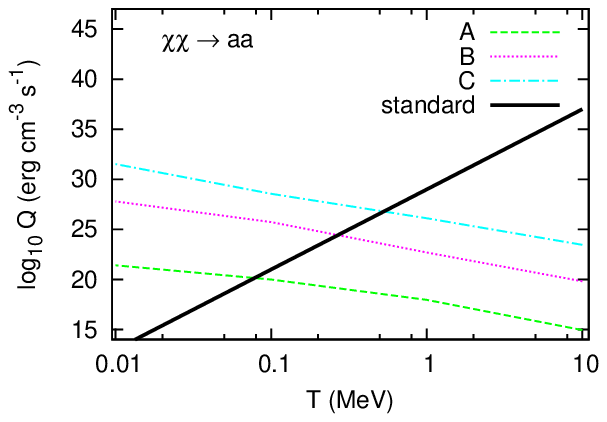}}
\caption{Energy emissivity from channels $\chi \chi \rightarrow \nu \nu$ (left) and via pseudoscalar mediators   $\chi \chi \rightarrow a a$ (right) with subsequent decay $ a \rightarrow \nu \nu$ as a function of temperature. Standard emission refers to MURCA processes. From \cite{cermeno3}.}
\label{Fig1}
\end{figure*}
Given the fact that inside the star the small DM fraction provides a Fermi energy $E_{F,\chi} \ll k_{\rm B}T$, in both reacttions, its distribution function can be approximated by a  classical Maxwell-Boltzmann type $f_\chi=f^{MB}_\chi(|\vec{p_i}|)=\left( \frac{1}{2\pi m_\chi k_B T}\right)^{\frac{3}{2}}n_\chi (r) e^{\frac{-|\vec{p_i}|^2}{2m_\chi T}}$. We can see that the phase space factor $f(f_1,f_2,f_3,f_4)$ in Eq. (\ref{qe}) introduces DM density and $T$ dependence into the calculation as a thermalized DM distribution exists inside the NS core. For the outgoing fermions the medium density effects have to be taken into account too through the phase space blocking factors as we have reviewed in Section 3, whereas for the outgoing pseudoscalars (bosons) no blocking has to be considered. 
\begin{table} \label{tab:models}
\centering
\begin{tabular}{|p{0.4cm}|p{1.5cm}|p{1.5cm}|p{1.5cm}|p{1.5cm}|}
\hline  & $m_\chi$ [GeV] & $m_a$ [GeV] & $g_{\chi}$ &  $g_0$  \\
\hline A & 0.1 & 0.05 & $7.5 \times 10^{-3}$ & $7.5 \times 10^{-3}$ \\
B & 1 & 0.05 & $1.2 \times 10^{-1}$ & $2 \times 10^{-3}$ \\ 
C & 30 & 1 & $6 \times 10^{-1}$ & $5 \times 10^{-5}$ \\
\hline
\end{tabular}
\caption{Parameters used in this work as appearing in the interaction lagrangian in Eq. (\ref{li}). $g_f=1$ is taking. From \cite{cermeno3}.}
\end{table}

In order to analyze an interesting astrophysical consequence, \cite{cermeno3} restrict the analysis to three different sets of flavour-universal ($g_f=1$) parameters that are not in conflict with existing phenomenology of direct detection experiments (\cite{bertone}) nor cosmological bounds (\cite{cosmo1, cosmo2}). The masses and couplings considered can be seen in Table I. Models A and B are mainly determined by DM relic abundance due to the fact that the DM mass is in the region where direct detection experiments are less restrictive but can be more constrained with Kaon decays. On the contrary, the couplings in set C are constrained by LUX results, \cite{lux}, in spin independent and spin dependent cross-sections. For more details see \cite{cermeno3}. On the other hand, they fixed the final fermion states to neutrinos. These weakly interacting SM fermions play a key role in releasing energy from NSs. At the end of the life of a very massive star the central core collapses gravitationally and it is well known that its gravitational binding energy is emitted in neutrinos (and antineutrinos) of the three families. A very efficient cooling scenario emerges in the first $\sim 10^5$ yr. Standard processes such as the URCA cooling or the modified URCA (MURCA) cooling, see \cite{friman, yako}, can release neutrinos with associated emissivities  $Q^{\rm URCA}_E\sim 10^{27} R (\frac{k_{\rm B}T}{0.1 \, \rm MeV})^6$ $\rm erg\, cm^{-3}\, s^{-1}$ and $Q^{\rm MURCA}_E\sim 10^{21} R (\frac{k_{\rm B}T}{0.1 \, \rm  MeV})^8$ $\rm erg \,cm^{-3} \,s^{-1}$, respectively. $R$ is a control function of order unity. These standard processes release energy from the baryonic system, having a very different effect in the energetic balance when compared to average energy injection from stellar DM annihilation processes, \cite{nsdm2}. It is important to remark that in this scenario it can be assumed neutrinos do not get trapped after being produced and therefore their Fermi-Dirac distributions fulfill $f_\nu \sim 0$.

In this context, to compare the result with the standard physics cooling, in the left panel of Fig. (\ref{Fig1}) the energy emissivity for the process $\chi \chi \rightarrow \nu \nu$ as a function of temperature for three sets of DM parameters in Table I is shown. The MURCA energy emissivity (solid line) is plotted too for the sake of comparison. Although this latter is not the only process contributing to the effective cooling, it sets an upper limit to standard emissivities in the scenario depicted due to the fact that to sustain the URCA fast reaction central densities must be higher to provide the $\sim 11\%$  protons fraction required (\cite{proton}). It is easy to see that around $T\sim 0.1$ MeV local central emissivities $\rm log_{10}\, (Q_E)\sim 21-22$ are balanced by the heating processes. It is important to remind ourselves that $T\sim 0.01 \; \rm MeV$ for a NS of $\sim 10^5$ yr. But the most important result is obtained for the reaction $\chi \chi \rightarrow a a$ with subsequent decay $a\rightarrow \nu \nu$, Fig.(\ref{Fig1}) right panel. In this case the neutrino emissivity is largely enhanced with respect to the direct production of neutrinos $\chi \chi \rightarrow \nu \nu$. For model C (somewhat heavier DM  particles) $Q_E$ matches and surpasses the standard MURCA emission at  $T\lesssim 1$ MeV while model B has analogous behavior for $T\lesssim 0.3$ MeV. Note that both models provide larger emissivities in this channel than the direct neutrino pair emission in the analyzed $T$ range in spite of being in a different range of masses. Model A provides similar results for both emission reactions, however.
This effect of enhanced emissivities can have an impact on internal temperatures and matter phases depending critically on them, as can be seen in \cite{stei}. In this sense, works as \cite{cas1,cas2} quote that the rapid cooling of the Cas A may be an indication of the existence of global neutron and proton superfluidity in the core.\\
As most of the fraction of DM particles will be in the central region of the NS, characterized by the length $r_{th}$, the impact of the radial extent of an emitting inner region through these new processes can be correlated to the ratio $\xi=(\sqrt{2} r_{\rm th}/R_b)$, which make reference to the fraction where $95\%$ of the approximately Gaussian distribution of DM particles can be found, versus  $R_b\sim 9$ km, radial extent of the boundary or limit of the core-to-crust region. Since the crust region has a tiny mass, in this scenario, it is not necessary to consider the refinement which it is taken into account in \cite{cermeno2}. For the models analyzed in \cite{cermeno3} this fraction ranges from $\xi \in[0.03,0.42]$ at $T\sim 1$ MeV to  $\xi \in[0.01,0.14]$ at $T\sim 0.1$ MeV. 

Focusing in a fixed stellar evolutionary time, $t$, to analyze the temperature evolution, the radial (internal) heat equation $r<R_b$ can be written as
\begin{equation}
\frac{2}{r}T'(r)+T^{''}(r)=\kappa^{-1}[Q_\nu-H(r)],
\label{heat}
\end{equation}
where $\kappa$ is the thermal conductivity given by a contribution from baryons and another from electrons \cite{yak} $\kappa=\kappa_b+\kappa_e$ and $Q_\nu$ is the standard MURCA emissivity, $Q_\nu\sim Q_E^{\rm MURCA}$. The DM heating term, $H(r)$, has the radial dependence induced by the DM density so that $H(r)=H_0e^{-(r/r_0)^2}$ with $H_0=Q_{E}^{aa}n^2_{0,\chi}$ and $r_0=r_{\rm th}(T)/\sqrt{2}$. Although a detailed solution of the full evolution equation remains for future works that should consider the dynamical part, in \cite{cermeno3} an study of the radial $T$-profile at different fixed evolutionary times is done solving Eq.(\ref{heat}) and assuming a flat initial profile to see how it is distorted from the presence of  internal cooling and heating terms. They find that, for example, fixing a value  $T(r)\equiv \bar{T}$ for $r\in [0, R_b]$ the central value $T_0$  is related to it through the approximate solution $T(r)=T_0+{\alpha} \frac{r^2}{6}-\beta \frac{r_0^2}{2} [1-\sqrt{\pi}\frac{r_0}{2} \frac{\rm erf(r/r_0)}{r}]$ where $\rm erf$ $(x)$ is the error function and ${\alpha}=Q_\nu (\bar{T})/\kappa({\bar{T}})$ and $\beta=H_0({\bar{T}})/\kappa ({\bar{T}})$ are coefficients that can be approximated at $\bar{T}$. Typical values can be seen in \cite{cermeno3}. The important thing here is that if we fixed, for example, $\bar{T}=10^9 K\sim 0.1\,\rm  MeV$ one can verify that the net effect of models B and C is heating the central volume while model A provides net cooling. However, it is important to remark that the star will self-adjust in a consisting way that has to be dynamically determined to see at what extent the subsequent thermal evolution is affected in the stellar volume. This could be studied through a full dynamical simulation to see how the cooling mechanisms adjust inside the star as a temporal sequence.

\section{CONCLUSIONS}
In summary, in this contribution we have reviewed some astrophysical consequences of the possible existence of dark matter by its interaction with ordinary matter ranging from low to high densities. In the parameter space of DM mass to cross-section off nucleons we have firstly considered fermionic DM particles in the light (low) mass region $m_\chi \lesssim 5$ GeV. We discuss the importance of in-medium effects, density and temperature, in DM-nucleon scattering when DM particles are gravitationally boosted into a dense compact object, where core densities typically exceed a few times  $n_{0}$ and temperatures $T\lesssim 50$ MeV.\\
Density effects have to be included through Fermi-Dirac distributions for the nucleon sector, which rectrict the parameter space, and effective values for the nucleon mass and its chemical potential, while temperature effects are present through the detailed balance factor, which takes into account the energy gained or lost due during the interaction. 
Temperature effects have been found to be important although to a lesser extent relative to density ones. The differential and integrated cross-sections are greatly affected by the finite density of matter, namely by the effect of a smaller  effective nucleon mass $m^*_N<m_N$. The mean free path for a DM particle is found to be larger than the typical values of those found for Standard Model neutrinos with vector-axial couplings. From these values obtained for the mean free path it can be concluded that the diffusive behavior approximation at finite density and temperature in the interior of NS is well grounded and DM can contribute to the energy transport in their interior. The simplified estimate for the mean free path, $\lambda_\chi\simeq 1/\sigma_{\chi N} n$, lacks the rich dependence on the phase space of the scattering process.
 
Then, restricting more the range of masses $m_\chi \lesssim 100 \;\rm MeV$, the interaction between DM particles and the periodically arrange of nuclei in the NS outer crust at lower ordinary densities can be studied. From this interaction phonons can be created in the lattice. Calculating the phonon excitation rate it can be seen that this rate is much larger than for cosmological neutrinos. As an astrophysical consequence of this new source of phonons the ion thermal conductivity in the outer envelope is modified founding that it can be largely enhanced at the highest LDM density $n_\chi\sim 100 n_{0,\chi}$ due to a net modification of the acoustic phonon population. Besides, comparing this result with electronic contribution to the conductivity for magnetized NSs in the perpendicular direction to the magnetic field an enhanced with respect to electron contribution in this direction can be found. This means that perpendicular thermal conductivity in these scenarios will be higher, and, although a detailed study of the quantitative effect in the surface temperature pattern remains to be undertaken, it is expected that this enhancement allows a reduction of the difference of heat transport among parallel and perpendicular directions to  the magnetid field. Based on previous works, as \cite{aniso1} and \cite{aniso2}, only including standard thermal contributions we expect that, as a natural consequence, the surface temperature profile would be more isotropic yielding flatter profiles for intermediate latitudes.

Another impact of this dark sector in stellar quantities of NSs is considered for a model of annihilating LDM. We use the Coy DM model, where the energy emissivity due to the annihilation of DM particles into neutrinos through a pseudoscalar interaction is taken into account. In this context and considering not only LDM particles it is shown that in the inner stellar regions where most of the DM population distribution exists the emissivity into neutrinos can be enhanced orders of magnitude with respect to the MURCA standard neutrino processes for parameter sets respecting constraints of direct detection limits and cosmological bounds. This novel mechanism of energy release may lead to enhanced cooling features. A detailed analysis is missing to determine whether locally warmer inner temperature stellar profiles that can affect the onset of superfluid phases at typical temperatures $T_c\sim 5\times 10^9$ K. Although some other heating processes have been quoted in the literature (rotochemical heating, \cite{chem}, or hot blobs located at different depths in the crust in young NS ,\cite{heater}) the qualitative picture arising from the DM annihilation process is different, as this heating in the inner region can produce thermal instabilities in the inner core. This could be an interesting field to explore in the future.

\begin{acknowledgements}

We thank useful discussions with R. Lineros, J. Edsjo and C. Albertus. This work has been supported by NewCompstar and MINECO Consolider-Ingenio { Multidark} CSD2009-00064 and FIS2015-65140-P projects. M. Cerme\~no is supported by a fellowship from the University of Salamanca.
\end{acknowledgements}

\bibliographystyle{pasa-mnras}
\bibliography{pasareview}

\end{document}